\documentclass[9pt,sigconf,screen]{acmart}
\usepackage{algorithm}
\usepackage{algorithmic}
\usepackage{amsmath}
\usepackage{xspace}
\usepackage{siunitx}
\usepackage{subfig}
\usepackage{tikz}
\AtBeginDocument{%
  \providecommand\BibTeX{{%
    \normalfont B\kern-0.5em{\scshape i\kern-0.25em b}\kern-0.8em\TeX}}}

\setcopyright{acmcopyright}
\copyrightyear{2018}
\acmYear{2018}
\acmDOI{XXXXXXX.XXXXXXX}

\acmConference[Arxiv]{ }{Nov. 2023}{Preprint}
\acmPrice{15.00}
\acmISBN{978-1-4503-XXXX-X/18/06}

\newcommand{\Prv}{$\mathcal{P}$\xspace}
\newcommand{\Vrf}{$\mathcal{V}$\xspace}
\newcommand{\Cir}{$\mathcal{C}$\xspace}
\newcommand\sys{$\mathsf{SPAM}$\xspace}

\begin{document}

\title{\textbf{SPAM}: Secure \& Private Aircraft Management}




\author{\textbf{Yaman Jandali*}, \textbf{Nojan Sheybani*}, \textbf{Farinaz Koushanfar} \\ 
University of California, San Diego \\ 
\tt\small{\{yeljanda, nsheybani, farinaz\}@ucsd.edu}}
\thanks{*Equal contribution}

\renewcommand{\shortauthors}{Jandali, Sheybani, and Koushanfar}

\begin{abstract}
With the rising use of aircrafts for operations ranging from disaster-relief to warfare, there is a growing risk of adversarial attacks. Malicious entities often only require the location of the aircraft for these attacks. Current satellite-aircraft communication and tracking protocols put aircrafts at risk if the satellite is compromised, due to computation being done in plaintext. In this work, we present \sys \footnote{The name \sys (Secure \& Private Aircraft Management) is crafted with a dual meaning in mind: signifying the management system's assurance of rendering externally observable aircraft data unintelligible, resembling typical ``spam'' content.}, a private, secure, and accurate system that allows satellites to efficiently manage and maintain tracking angles for aircraft fleets without learning aircrafts' locations. \sys is built upon multi-party computation and zero-knowledge proofs to guarantee privacy and high efficiency. While catered towards aircrafts, \sys's zero-knowledge fleet management can be easily extended to the IoT, with very little overhead.
\end{abstract}

\ccsdesc[500]{Security and privacy~Cryptography}
\ccsdesc[300]{Computer systems organization~Embedded and cyber-physical systems}
\ccsdesc[300]{Computer systems organization~Real-time systems}

\keywords{Privacy-Preserving Computation, Zero-Knowledge Proofs, Aircraft Privacy, Multi-Party Computation, Two-Party computation}



\received{20 February 2007}
\received[revised]{12 March 2009}
\received[accepted]{5 June 2009}

\maketitle

\begin{figure}[h]
  \centering
  \includegraphics[width=0.8\columnwidth]{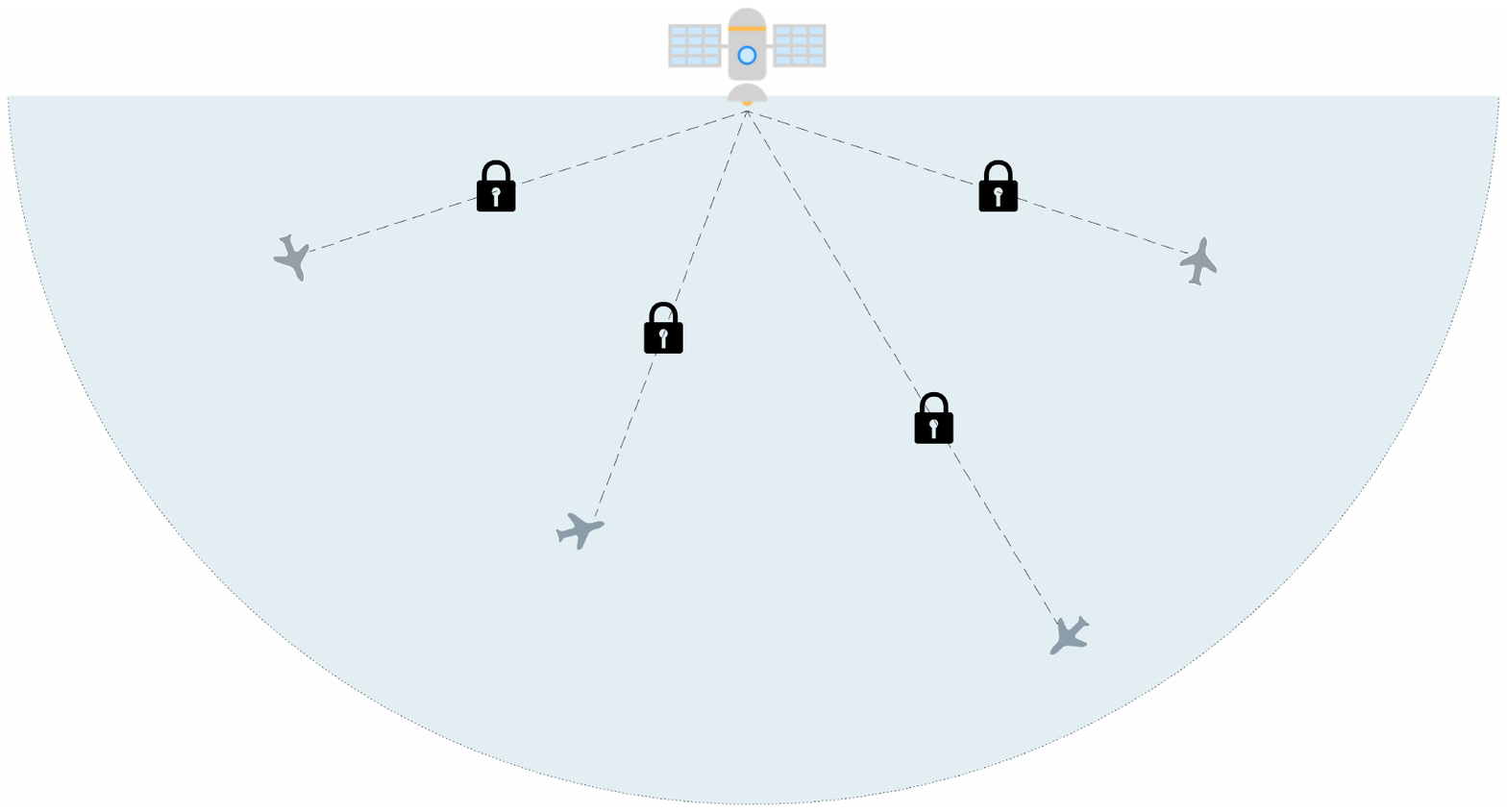}
  \caption{\sys is an end-to-end aircraft management system that guarantees location privacy to both the aircraft and satellite via scalable Boolean synthesis for secure 2PC and ZKP.}
\end{figure}

\section{Introduction}
Over the past decade, the use of piloted aircrafts, as well as Unmanned Aerial Vehicles (UAVs), has significantly grown. Fleets of remotely controlled vehicles are increasingly used in disaster relief, search and rescue, warfare, and more \cite{disRelief, searchAndRescue}. The Federal Aviation Administration (FAA) of the United States of America (USA) handles approximately 16.5 million flights each year. In addition to the aircrafts, the USA’s Department of Defense operates over 11,000 Unmanned Aircraft Systems (UAS) \cite{dod}. This includes UAVs, commonly known as "drones". With the increasing number of aircrafts, not just in the USA, but globally, there is also an increased risk of one of these aircrafts  becoming the target of an adversarial attack or of “going rogue” (i.e. falling into the hands of enemy forces \cite{rogueNations}), and there have already been cases where this is suspected to have happened \cite{hartmann}. 

Many types of adversarial attacks, such as those using automated weaponry based on GPS tracking, GPS spoofing, jamming attacks, and more, rely on knowledge of an aircraft's location \cite{GPSNav}. 
In addition, satellite-based control systems that require an antenna to be aimed at a remotely controlled aircraft, as well as satellites that provide other types of aircrafts with communication capabilities, can have their transmissions be intercepted \cite{interceptComm}, thereby revealing the location of the tracked vehicle. For this reason, it is imperative that the computations used to determine the direction that a satellite's antenna faces be done without ultimately revealing the location of the aircraft. This can be accomplished by taking advantage of scalable methods for privacy preserving computation based on synthesis of Boolean logic \cite{tg}.


Works such as \cite{PPCA, hideAndSeek} offer solutions aiming to protect the privacy of a drone's location. However, this is done either by simply communicating a temporary location through a secure channel or by obfuscating the location through differential privacy, which would make it infeasible to accurately compute the trajectory from one party to the other. Akkaya et al. \cite{Akkaya} focus on private communication of data from an aircraft to a server for machine learning inference on private data, but do not address concerns for privacy of the aircraft's location. Other methods such as \cite{blockchain3} utilize blockchain for secure authentication of drones, but this approach is ineffective for privacy-preserving location management. 

Challenges of privacy preservation are further amplified by the communication channels available between aircrafts and satellite. In this work, we assume that aircrafts and satellites communicate using the standardized SATCOM network. SATCOM enables fast aircraft-satellite communication, but lacks encryption below the application layer by default. This puts any communication through SATCOM at risk of eavesdropping attacks, which has been drastically simplified due to the introduction of software defined radios (SDRs) \cite{9811060}. The solutions require different communication infrastructures or the adoption of secure channels, which increases the overhead and limits the scalability. Scalability is essential, as in many satellite-aircraft communication settings, satellites are actively communicating with many aircrafts at a time for various tasks. One of the most prominent tasks in this domain is the enforcement of certain routes or bounded areas for an aircraft to traverse (e.g. surveillance aircrafts). It is essential to ensure aircraft location privacy in these settings, as eavesdroppers and malicious adversaries that compromise the satellite could mount dangerous attacks if this location is leaked.

In this work, we propose \sys, the first automated end-to-end solution to private satellite-aircraft localization. \sys utilizes automated Boolean logic synthesis to build optimal circuits for privacy preserving functions to preserve the secrecy of both aircraft and satellite location. Using a combination of Boolean logic-based two-party computation and zero knowledge proofs, \sys enables two key tasks in satellite-aircraft localization with strong privacy guarantees: 
( 1 ) \textit{Movement Tracking}: \sys enables a satellite to continuously adjust its antenna to maximize communication efficiency without revealing the satellite or aircraft's location and ( 2 ) \textit{Location Management}: \sys enables a satellite to monitor aircrafts within its network and ensure that they are staying within certain bounds, without learning the exact locations of the aircrafts.

To the best of our knowledge, we propose the first method that securely verifies the correct angle of a satellite's antenna for aircraft communication. Further, \sys is the first work to allow satellites to enforce location bounds for aircrafts in their networks, without revealing the exact location of the aircrafts. \sys not only elegantly solves these problems with very little overhead, but also defends against eavesdropping attacks on SATCOM networks due to the use of privacy-preserving computation.





In summary, our contributions are as follows.
\begin{itemize}
    \item Introduction of \sys, the first automated end-to-end framework for privately tracking and managing aircrafts, without revealing any location information.
    \item Enabling unintelligible tracking of location (without recompute) by introducing new 2PC-based methods for further private tracking of aircraft trajectories.
    \item Utilizing scalable Boolean logic synthesis to design novel circuits compatible with interactive zero-knowledge proof protocols which enforces spatial bounds for aircrafts.
    \item Extensive evaluations of \sys highlight its performance at scale, while providing privacy to all parties involved.
\end{itemize}

\section{Preliminaries}\label{prelim}

With increasing data-driven decision making and information sharing, there has been a growing need for privacy preserving computation to protect our sensitive data. A variety of solutions have emerged over the years, with one efficient and robust choice being multi-party computation (MPC), which allows for joint computation on private data by multiple parties while achieving provable privacy and accuracy.

\subsection{Multi-Party Computation}\label{MPCSec}

In MPC schemes, $n$ parties, each holding their own private inputs denoted as ${d_1, d_2, ..., d_n}$, perform joint computation of a public function, $F(d_1, d_2, ..., d_n)$, such that only the output of this function is revealed while the each party's input remains private. There are generally two settings of MPC -  namely \textit{semi-honest} and \textit{malicious} settings. In semi-honest settings, parties may gather information passively without straying from the designated protocol. In malicious settings, parties are able to actively pursue information both passively, as with the semi-honest setting, but also by deviating from the designated protocol to gain more information \cite{introToMPC}. 
In this work, we utlize two-party computation, a subset of multi-party computation.
\subsubsection{Two-Party Computation}
Two-party computation (2PC) was first introduced by Andrew Yao in 1986 and was later extended by Goldreich, Micali, and Wigderson to MPC \cite{2pc, mpc}. 2PC often relies on the use of garbled circuits, a topic first introduced by Yao but further formalized by Beaver et al. \cite{BMR}. Garbled circuit based evaluation involves two parties, namely a \textit{garbler} and an \textit{evaluator}. The garbler's role is in generating a circuit that describes an underlying function to be computed with both parties' inputs. The evaluator receives this circuit from the garbler and by using Oblivious Transfer \cite{OT}, the evaluator is able to garble its own input with the help of the garbler, without the garbler learning the evaluator's input. The function is then evaluated, which results in both parties obtaining the function's output without revealing any information about each other's inputs in addition to what can be inferred from the output itself. Beaver-Micali-Rogaway (BMR) is one method of implementing garbled circuits by utilizing generalized secure protocols for 2PC to compute the garbled circuit \cite{BMR}. 


\subsection{Zero-Knowledge Proofs}
Zero-Knowledge Proofs (ZKPs) are a two-party cryptographic primitive between two parties: prover \Prv and verifier \Vrf. ZKPs enables \Prv to prove that the the evaluation of a computation \Cir on a private value $w$, called the witness, is valid without revealing anything about $w$. In standard ZKP schemes, \Prv convinces \Vrf that $w$ is a valid input such that $y=\mathcal{C}(x,w)$, where $x$ and $y$ are public inputs and outputs, respectively \cite{zk}.
Interactive ZKP schemes require many rounds of interaction between \Prv and \Vrf to successfully build a ZKP. Non-interactive ZKPs (NIZKs) allow for proof generation to be done in one step, but often require a rigorous trusted setup process per computation \Cir. While NIZKs have smaller proof size and faster verification, they benefit from \textit{publicly-verifiable} proofs, meaning that any \Vrf can verify the proof. 

Conversely, interactive ZKPs have larger proofs and are \textit{designated verifier}, meaning that only the verifier interacting with \Prv can verify the proof, however do not have a trusted setup process. One of the main advantages of interactive proofs is the reduced prover complexity when compared to NIZKs. This becomes especially suitable when working with resource-constrained devices, such as those in IoT settings, which we discuss in detail in section \ref{eval}. While the \textit{designated verifier} constraint can viewed as a drawback, we will explain that this is actually suitable for our system, and far outweighs the drawback of having a rigorous trusted setup process. In this work, we employ the Wolverine protocol \cite{passwordAuth}, a state-of-the-art interactive ZKP system that is notably efficient in terms of execution time, memory demands, and data transfer for \Prv. Within this protocol, the witness $w$ is verified using information-theoretic message authentication codes (IT-MACs). Computations in Wolverine are structured as either arithmetic or Boolean circuits, and are collaboratively evaluated by \Prv and \Vrf. Upon completion, \Prv reveals the output to demonstrate the validity of the proof, which in our case is proving that the evaluated output matches a pre-existing public output. 
 
\section{Related Work}\label{relatedWork}

Works such as PPCA \cite{PPCA} recognize the threats of location leakage to adversaries. With increasing investments from stakeholders such as Google and Amazon, there are more drones being used for autonomous delivery each year. The authors claim to be the first ones to develop an improved collision detection system that maintains the privacy of the drone locations rather than having each UAV openly broadcasting its location - the current norm. They do so by having UAVs share temporary locations through secure wireless connections to nearby drones that have a risk of colliding with each other. 

In Hide and Seek \cite{hideAndSeek}, the authors show that Critical Infrastructure operators are able to preserve drone privacy while detecting drones violating no-fly zone designations. They provide DiPrID, a framework which uses differential privacy to improve drone location privacy. They also propose ICARUS, which provides a solution for detecting unauthorized drones within no-fly zones. 

The authors of \cite{IODdesign} outline the lack of mechanisms for the privacy preservation of drone locations but do not provide implementation to fill this gap. \cite{jammingAttacks, attacks2} further demonstrate the need for location privacy by showcasing various attacks on drones, such as jamming and GPS spoofing, that rely on knowing a drone's location.

Works such as \cite{Akkaya, darksky} discuss the growing issues of drones capturing sensitive data and provide frameworks for enabling privacy-preserving transmission of data captured by drones, such as video recordings. However, they do not discuss the privacy of the drone itself. 

Research interest in efficient privacy preserving computation has drastically increased in recent years. Developments in the field have allowed for advancements in a variety of private computing applications such as machine learning model inference on sensitive data, credit scoring, identity verification, and more \cite{passwordAuth, chameleon, creditScoring}. A number of methods are used, such as multiparty computation and homomorphic encryption, which allow for secure computation on private data. Alongside this, zero-knowledge proofs are an emerging technology that allow users to prove attributes about their data, without revealing anything about their data \cite{zk}.

\section{Methodology}\label{methodology}


\begin{figure*}[t]
\centering
\subfloat[Trajectory Computations: The aircraft and satellite jointly compute the unit vector describing the desired trajectory for the satellite's antenna using Two-Party Computation.]{\includegraphics[width=.4\textwidth]{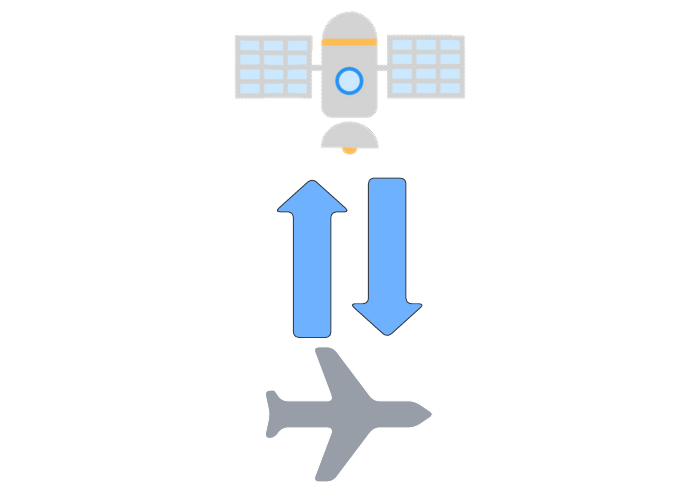}}
\hspace{1em}
\subfloat[Range Checks: Aircrafts periodically send a ZKP to prove to the satellite that they are still within range.]{\includegraphics[width=.4\textwidth]{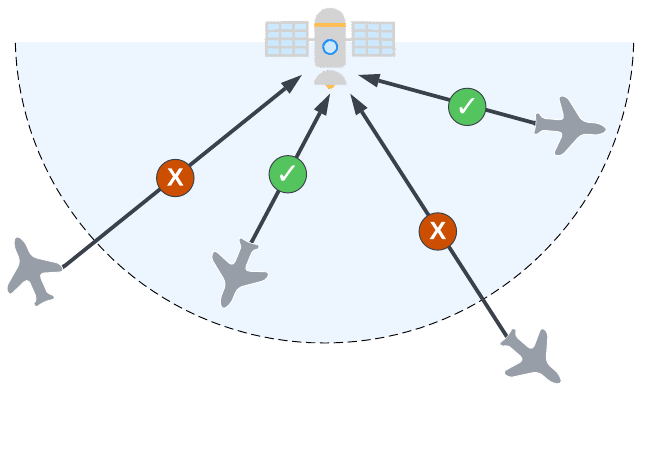}}
\caption{High Level Overview of \sys.}
\label{fig:system}
\end{figure*}

Many aircrafts, including UAS or drones, are often remotely controlled through satellite systems, which aim an antenna towards them  to achieve efficient transmission. Naively, this can be accomplished through the aircraft communicating its location to the satellite.  However,  this approach exposes the aircraft's location to the satellite, introducing vulnerabilities  if the satellite is compromised by adversaries. 

\sys provides secure computation of the trajectory from a spacecraft to an aircraft through 2PC and a fleet management system built on zero-knowledge proofs, all without compromising the location of either party. We represent the trajectory using a \textit{trajectory unit vector}, which is simply a vector of magnitude one that defines the direction from the satellite to the aircraft in three dimensional space.


\subsection{Threat Model(s)}

\sys aims to protect the privacy of the aircraft's location from the satellite, as leaking this could make the aircraft susceptible to attacks from adversaries. Alongside this, we would like to leak as little information as possible about the satellite to the aircraft, which is why two-party computation and zero knowledge proofs are employed in \sys. In our setting, we assume that both parties are \textit{semi-honest}, meaning that they will follow the given protocol, but try to learn as much information as possible with the data in hand. While not the main contribution of this work, we also consider the threat of a malicious satellite that may deviate from the protocol.

\subsection{\sys Overview}
\sys consists of two core functionalities: movement tracking and location management.
The main task of \sys's movement tracking is privately updating the angle of a satellite's antenna to ensure that it is pointing at the aircraft, without revealing the exact location of the aircraft. The main task of \sys's location management is to generate and verify proofs in which aircrafts attest to the satellite that they are bound-compliant. We note that we use scalable Boolean logic synthesis \cite{emp, tg} to optimize our 2PC and ZKP circuits, ensuring state-of-the-art performance.

\subsubsection{Movement Tracking}
An aircraft or drone is equipped with a communication system that requires a satellite connection. In order to transmit information most efficiently between the aircraft and the satellite, an antenna on the satellite must point itself at the aircraft's changing location. We denote the aircraft by $p$, the satellite by $s$, and the locations of the satellite and the aircraft by $(x_s, y_s, z_s)$  and $(x_a, y_a, z_a)$, respectively. If the satellite's location is public,  we are able to simply have the aircraft perform the computation of the satellite's trajectory unit vector and send the computed vector to the satellite. This fulfills all of privacy requirements for the aircraft without the need for expensive private computation. This simple algorithm is described in Algorithm \ref{algo:oneWayAlg}.




\begin{algorithm}
\caption{Plain-text Trajectory Unit Vector Calculation}
\label{algo:oneWayAlg}

\begin{algorithmic}
\STATE \textbf{Input}:
\STATE Satellite location (sent to aircraft): $(x_s, y_s, z_s)$
\STATE aircraft location (known by aircraft): $(x_a, y_a, z_a)$
\STATE \textbf{Computed}:
\STATE Vector $v = (x_v=x_a - x_s, y_v=y_a - y_s, z_v=z_a - z_s)$
\STATE Magnitude $m = \sqrt{x_v^2 + y_v^2 + z_v^2}$
\STATE Unit vector $a = \left(\frac{x_v}{m}, \frac{y_v}{m}, \frac{z_v}{m}\right)$
\STATE \textbf{Send the Unit vector to the satellite.}
\end{algorithmic}
\end{algorithm}



However, under \sys's stricter privacy constraints, we assume that both of the aircraft's and satellite's locations must remain private.  We rely on 2PC to securely compute the trajectory. In this more secure setting, we perform the same computations as we have in Algorithm \ref{algo:oneWayAlg} but using secure protocols. This new formulation is described in Algorithm \ref{algo:twoWayAlg}. We build our computation using semi-honest and malicious secure two-party methods based on oblivious transfer (OT). For the semi-honest threat model, we utilize the Semi \cite{mpspdz} protocol, which utilizes OT and secret sharing over a large prime field to support semi-honest computation. For the malicious threat model, we use the MASCOT \cite{mascot} protocol, which builds off of Semi. MASCOT primarily adds MAC keys and OT correlation and consistency checks to Semi to support malicious parties.
The utilization of these methods enable \sys to securely compute the trajectory unit vector with the aircraft's and satellite's locations as private inputs. After secure computation, the resulting trajectory unit vector is only revealed to the satellite. This is done to protect the satellite's location, as the aircraft could triangulate the satellite's location if more than one unit trajectory vector was learned during movement.

\begin{algorithm}
\caption{Private Trajectory Unit Vector Calculation}
\label{algo:twoWayAlg}

\begin{algorithmic}
\STATE Satellite location: $(x_s, y_s, z_s)$
\STATE Aircraft location: $(x_a, y_a, z_a)$
\STATE $Party 1 \leftarrow Satellite$
\STATE $Party 2 \leftarrow aircraft$
\STATE \textbf{Computed}:
\STATE $f$ $\leftarrow$ $gen\_garble\_circuit(get\_trajectory(a, b))$
\STATE $g_1 \leftarrow{garble} (x_s, y_s, z_s)$
\STATE $g_2 \leftarrow{garble} (x_a, y_a, z_a)$
\STATE $reveal\_to\_party\_1(f(g_1, g_2))$
\STATE \textbf{Note: get\_trajectory( ) is the function described by Algorithm \ref{algo:oneWayAlg}}
\end{algorithmic}

\end{algorithm}

\subsubsection{Location Management}
Satellites are often in communication with many aircrafts at a given time. In tasks, such as surveillance, a satellite would like to enforce that the aircrafts are staying within these bounds. However, the aircraft wants to ensure that their location within these bounds is not revealed, in the event of the satellite being compromised. \sys enables this with a provably private location management scheme. \sys's location management is built upon ZKPs, which enables the aircrafts to prove that they are within bounds that are predetermined and updated by the satellite, while ensuring that the specific locations of the aircrafts are not revealed in the event of the satellite being compromised.

The bounds that the satellite enforces are established in the form $[(x_{min}, x_{max}), (y_{min}, y_{max}), (z_{min}, z_{max})]$. We design an efficient Boolean circuit for the $\geq$ and $\leq$ comparisons, based on fast custom multiplexers, that is compatible with the Wolverine protocol that returns a single bit. These circuits are used to compare the aircraft's current location $(x_a, y_a, z_a)$ with the satellites' enforced bounds in zero-knowledge. As computation in Wolverine is represented as Boolean circuits, we aim to take advantage of this by performing primarily bitwise operations. The high-level algorithm we employ can be see in algorithm \ref{algo:location}. As Wolverine has optimized implementations of zero-knowledge AND gates, \sys's location management requires very little overhead.

\begin{algorithm}
\caption{Check if aircraft is in bounds}
\label{algo:location}
\begin{algorithmic}
\STATE Aircraft $\gets$ Prover\\
\STATE Satellite $\gets$ Verifier\\
\STATE \textbf{Input}:
\STATE Bounds: $[(x_{min}, x_{max}), (y_{min}, y_{max}), (z_{min}, z_{max})]$
\STATE Aircraft location (known by aircraft): $(x_a, y_a, z_a)$
\STATE \textbf{ZK Circuit}:
\STATE Bit $x$ = $x_a.geq(x_{min})$ AND $x_{a}.leq(x_{max})$
\STATE Bit $y$ = $y_a.geq(y_{min})$ AND $y_{a}.leq(y_{max})$
\STATE Bit $z$ = $z_a.geq(z_{min})$ AND $z_{a}.leq(z_{max})$
\STATE Bit $valid$ = $x$ AND $y$ AND $z$
\STATE \textbf{Return $valid$}
\end{algorithmic}

\end{algorithm}


\subsection{Implementation Details}

While there are many coordinate systems for representing location in 3D space, we chose to evaluate on the Cartesian coordinate system. Other methods, such as the use of polar coordinates, often rely on trigonometric functions to determine the distance and direction between points in space. These functions are often more expensive to compute, particularly in the realm of privacy-preserving computing. 

\sys's 2PC location tracking implementation is built using \textit{MP-SPDZ} \cite{mpspdz}. MP-SPDZ is a user-friendly library that allows for implementation of 2PC while leveraging a variety of protocols to use in the backend. This enables benchmarking of different contexts, such as in semi-honest and malicious settings. We use  \sys's zero-knowledge location management is built using the Wolverine \cite{weng2021wolverine} protocol, an efficient interactive ZKP protocol that boasts highly scalable communication and performance. For both applications, we utilize state-of-the-art scalable Boolean logic synthesis tools \cite{emp, tg} to design low-level, optimized representations of our computation.




\section{Evaluation and Discussion}\label{eval}
\subsection{Setup}
The end-to-end \sys framework is implemented in C++ and the MP-SPDZ domain specific language.
We use the EMP-Toolkit~\cite{emp} for implementation of zero-knowledge proofs.
We run all experiments on a 128GB RAM, AMD Ryzen 3990X CPU desktop.
\subsection{Results}

We benchmark \sys's capabilities over several bitwidths ranging from 128 to 8 bits. We note that we do not perform separate experiments for the semi-honest and malicious threat model for the location management task as Wolverine is designed to be maliciously secure, however supports semi-honest parties. The results can be seen in Tables \ref{tab:movement} and \ref{tab:manage}. As can be seen in both applications, the communication is not affected by the bitwidths. This is due to the fact that all the specific protocols - Semi, MASCOT, and Wolverine - that are used in \sys are designed such that communication grows with respect to the size of the computation/circuit. Due to the lean, optimized algorithms we design for \sys, the communication cost is negligible in the semi-honest movement tracking and location management tasks. We see a significant increase in communication cost and runtime in the malicious location management setting, as MASCOT introduce OT correlation checks and MAC generation schemes to ensure that the protocol is completely soundly.

\begin{table}[hbt!]
  \centering
  \small
  \resizebox{\columnwidth}{!}{\begin{tabular}{ccccc}
    \toprule
    & \multicolumn{2}{c}{Semi-Honest} & \multicolumn{2}{c}{Malicious} \\
    \cmidrule(lr){2-3} \cmidrule(lr){4-5}
    \textbf{Bitwidth} & \multicolumn{1}{c}{\textbf{Runtime (s)}} & \multicolumn{1}{c}{\textbf{Comm. (MB)}} & \multicolumn{1}{c}{\textbf{Runtime (s)}} & \multicolumn{1}{c}{\textbf{Comm. (MB)}} \\
    \midrule
     128 & 0.13 & 0.75 & 3.44 & 168.40 \\
    100 & 0.09 & 0.75 & 3.18 & 168.40 \\
     64 & 0.08 & 0.75 & 3.13 & 168.40 \\
     32 & 0.08 & 0.75 & 3.12 & 168.40 \\
     16 & 0.08 & 0.75 & 2.96 & 168.40 \\
     8 & 0.06 & 0.75 & 2.95 & 168.40 \\
    \bottomrule
  \end{tabular}}
\caption{Evaluation of \sys's Movement Tracking}
  \label{tab:movement}
\end{table}


\begin{table}[hbt]
    \centering
    \resizebox{\columnwidth}{!}{\begin{tabular}{ccccc}
    \toprule
    & \multicolumn{2}{c}{Prover (\Prv)} & \multicolumn{2}{c}{Verifier (\Vrf)} \\
    \cmidrule(lr){2-3} \cmidrule(lr){4-5}
     \textbf{Bitwidth} & \textbf{Runtime (ms)} & \textbf{Comm. (kB)} & \textbf{Runtime (ms)} & \textbf{Comm. (kB)} \\
     \midrule
     128 & 53.87 & 84.54 & 44.88 & 105.12\\ 
     100 & 33.39 & 84.54 & 34.45 & 105.12\\
     64 & 24.79 & 84.54 & 23.32 & 105.12\\
     32 & 11.5 & 84.54 & 10.14 & 105.12\\
     16 & 6.71 & 84.54 & 6.92 & 105.12\\
     8 & 5.96 & 84.54 & 6.08 & 105.12\\
    \bottomrule
    \end{tabular}}
    \caption{Evaluation of \sys's Location Management}
    \label{tab:manage}
\end{table}

In semi-honest movement tracking and location management, \sys boasts very low runtime, making \sys a real-time solution for private aircraft management. While the malicious movement tracking setting leads to much higher runtime, this is a common pitfall of the malicious threat model. Due to the significantly increased amount of computation that needs to be done to ensure that a malicious party does not break the protocol, this increased runtime is unavoidable. We provide the measurements of runtime for several bitwidths, from 128 bits to 8 bits. This is done to show how \sys overhead changes based on the amount of precision that is necessary to successfully locate or prove the location of an aircraft. Based on the bandwidth and computational power that is available at a given time, the bitwidth can be adjusted to reduce the computational overhead or satellite-aircraft interaction time. 

The presented results highlight the communication and computational efficiency that \sys achieves by utilizing novel techniques combined with state-of-the-art privacy-preserving protocols. Unlike most privacy-preserving approaches, \sys is not limited by scale, as the circuit sizes and the number of parties per computation stay constant. The only minor bottleneck is communication, which only becomes a problem if the network, SATCOM in our case, has limited bandwidth. Overall, the results show that \sys is a real-time solution to private aircraft management that can scale to real-world applications.

\subsection{Extending \sys}
As a brief aside, we want to discuss the extension of \sys into the IoT - particularly the zero-knowledge location management aspect. Oftentimes, edge devices have computational and bandwidth constraints, which makes it challenging to implement privacy-preserving solutions in the IoT. Due to the lightweight communication and computational overhead that we show for \sys, we believe that this work can be easily extended to the IoT and provide a secure solution towards edge devices proving attributes of their collected data. 
In \sys, we have aircrafts proving that their current location is within certain pre-determined bounds using ZKP techniques.  There are many works that demonstrate the feasibility of ZKP-based systems for IoT devices \cite{iot1,litezkp, iot2}. It is therefore expected that the ZKP techniques employed in \sys can be extended to different sensor-based IoTs. For instance, users that utilize biomedical IoT edge devices can adopt the \sys approach to prove that their health vitals (e.g. heart rate) are within a certain "safe" range. As health data is often sensitive, the user would prefer that this data stays private, unless their vitals are deemed to be in an unsafe range. While this is just an example, it is clear to see that \sys's approach towards aircraft location management can be modified to span many domains within the IoT. 

\section{Conclusion}
In this work, we presented \sys, a privacy preserving framework for optimizing transmission efficiency and aircraft management. By leveraging methods for secure two-party computation, \sys preserves the privacy of each party's location while enabling a satellite to maximize its communication throughput by securely computing the optimal trajectory for its antennas to take. \sys also utilizes zero-knowledge proof techniques to provide a monitoring system that enables a satellite to determine if an aircraft's location falls within a predetermined range without revealing the exact location of the aircraft. Automated Boolean logic synthesis is utilized for representing computation to ensure state-of-the-art runtime and communication for our given application. While many standard methods of satellite communication provide little privacy preservation, \sys implements an automated end-to-end system for protection against both semi-honest and malicious adversaries, as well as security against eavesdropping attacks, with little overhead. 


\bibliographystyle{ACM-Reference-Format}
\bibliography{sample-base}


\begin{thebibliography}{33}


\ifx \showCODEN    \undefined \def \showCODEN     #1{\unskip}     \fi
\ifx \showDOI      \undefined \def \showDOI       #1{#1}\fi
\ifx \showISBNx    \undefined \def \showISBNx     #1{\unskip}     \fi
\ifx \showISBNxiii \undefined \def \showISBNxiii  #1{\unskip}     \fi
\ifx \showISSN     \undefined \def \showISSN      #1{\unskip}     \fi
\ifx \showLCCN     \undefined \def \showLCCN      #1{\unskip}     \fi
\ifx \shownote     \undefined \def \shownote      #1{#1}          \fi
\ifx \showarticletitle \undefined \def \showarticletitle #1{#1}   \fi
\ifx \showURL      \undefined \def \showURL       {\relax}        \fi
\providecommand\bibfield[2]{#2}
\providecommand\bibinfo[2]{#2}
\providecommand\natexlab[1]{#1}
\providecommand\showeprint[2][]{arXiv:#2}

\bibitem[dod({[n.\,d.]})]%
        {dod}
 \bibinfo{year}{[n.\,d.]}\natexlab{}.
\newblock \bibinfo{title}{{UNMANNED AIRCRAFT SYSTEMS (UAS)} DoD Purpose and Operational Use}.
\newblock \bibinfo{howpublished}{\url{https://dod.defense.gov/UAS/}}.
\newblock
\newblock
\shownote{Accessed: 2023-11-02}.


\bibitem[Akkaya et~al\mbox{.}(2019)]%
        {Akkaya}
\bibfield{author}{\bibinfo{person}{Kemal Akkaya}, \bibinfo{person}{Vashish Baboolal}, \bibinfo{person}{Nico Saputro}, \bibinfo{person}{Selcuk Uluagac}, {and} \bibinfo{person}{Hamid Menouar}.} \bibinfo{year}{2019}\natexlab{}.
\newblock \showarticletitle{Privacy-Preserving Control of Video Transmissions for Drone-based Intelligent Transportation Systems}. In \bibinfo{booktitle}{\emph{2019 IEEE Conference on Communications and Network Security (CNS)}}. \bibinfo{pages}{1--7}.
\newblock
\urldef\tempurl%
\url{https://doi.org/10.1109/CNS.2019.8802665}
\showDOI{\tempurl}


\bibitem[Arteaga et~al\mbox{.}(2019)]%
        {attacks2}
\bibfield{author}{\bibinfo{person}{Sandra~P{\'e}rez Arteaga}, \bibinfo{person}{Luis Alberto~Mart{\'\i}nez Hern{\'a}ndez}, \bibinfo{person}{Gabriel~S{\'a}nchez P{\'e}rez}, \bibinfo{person}{Ana Lucila~Sandoval Orozco}, {and} \bibinfo{person}{Luis Javier~Garc{\'\i}a Villalba}.} \bibinfo{year}{2019}\natexlab{}.
\newblock \showarticletitle{Analysis of the GPS spoofing vulnerability in the drone 3DR solo}.
\newblock \bibinfo{journal}{\emph{IEEE Access}}  \bibinfo{volume}{7} (\bibinfo{year}{2019}), \bibinfo{pages}{51782--51789}.
\newblock


\bibitem[Baselt et~al\mbox{.}(2022)]%
        {9811060}
\bibfield{author}{\bibinfo{person}{Georg Baselt}, \bibinfo{person}{Martin Strohmeier}, \bibinfo{person}{James Pavur}, \bibinfo{person}{Vincent Lenders}, {and} \bibinfo{person}{Ivan Martinovic}.} \bibinfo{year}{2022}\natexlab{}.
\newblock \showarticletitle{Security and Privacy Issues of Satellite Communication in the Avlatlon Domain}. In \bibinfo{booktitle}{\emph{2022 14th International Conference on Cyber Conflict: Keep Moving! (CyCon)}}, Vol.~\bibinfo{volume}{700}. \bibinfo{pages}{285--307}.
\newblock
\urldef\tempurl%
\url{https://doi.org/10.23919/CyCon55549.2022.9811060}
\showDOI{\tempurl}


\bibitem[Beaver et~al\mbox{.}(1990)]%
        {BMR}
\bibfield{author}{\bibinfo{person}{Donald Beaver}, \bibinfo{person}{Silvio Micali}, {and} \bibinfo{person}{Phillip Rogaway}.} \bibinfo{year}{1990}\natexlab{}.
\newblock \showarticletitle{The round complexity of secure protocols}. In \bibinfo{booktitle}{\emph{Proceedings of the twenty-second annual ACM symposium on Theory of computing}}. \bibinfo{pages}{503--513}.
\newblock


\bibitem[Benjamin(2013)]%
        {rogueNations}
\bibfield{author}{\bibinfo{person}{Medea Benjamin}.} \bibinfo{year}{2013}\natexlab{}.
\newblock \bibinfo{booktitle}{\emph{Drone warfare: Killing by remote control}}.
\newblock \bibinfo{publisher}{Verso Books}.
\newblock


\bibitem[Boo et~al\mbox{.}(2021)]%
        {litezkp}
\bibfield{author}{\bibinfo{person}{EunSeong Boo}, \bibinfo{person}{Joongheon Kim}, {and} \bibinfo{person}{JeongGil Ko}.} \bibinfo{year}{2021}\natexlab{}.
\newblock \showarticletitle{LiteZKP: Lightening zero-knowledge proof-based blockchains for IoT and edge platforms}.
\newblock \bibinfo{journal}{\emph{IEEE Systems Journal}} \bibinfo{volume}{16}, \bibinfo{number}{1} (\bibinfo{year}{2021}), \bibinfo{pages}{112--123}.
\newblock


\bibitem[Brighente et~al\mbox{.}(2022)]%
        {hideAndSeek}
\bibfield{author}{\bibinfo{person}{Alessandro Brighente}, \bibinfo{person}{Mauro Conti}, {and} \bibinfo{person}{Savio Sciancalepore}.} \bibinfo{year}{2022}\natexlab{}.
\newblock \showarticletitle{Hide and Seek: Privacy-Preserving and FAA-Compliant Drones Location Tracing}. In \bibinfo{booktitle}{\emph{Proceedings of the 17th International Conference on Availability, Reliability and Security}} (Vienna, Austria) \emph{(\bibinfo{series}{ARES '22})}. \bibinfo{publisher}{Association for Computing Machinery}, \bibinfo{address}{New York, NY, USA}, Article \bibinfo{articleno}{134}, \bibinfo{numpages}{11}~pages.
\newblock
\showISBNx{9781450396707}
\urldef\tempurl%
\url{https://doi.org/10.1145/3538969.3543784}
\showDOI{\tempurl}


\bibitem[Chinthi-Reddy et~al\mbox{.}(2022)]%
        {darksky}
\bibfield{author}{\bibinfo{person}{Samhith~Reddy Chinthi-Reddy}, \bibinfo{person}{Sunho Lim}, \bibinfo{person}{Gyu~Sang Choi}, \bibinfo{person}{Jinseok Chae}, {and} \bibinfo{person}{Cong Pu}.} \bibinfo{year}{2022}\natexlab{}.
\newblock \showarticletitle{DarkSky: Privacy-preserving target tracking strategies using a flying drone}.
\newblock \bibinfo{journal}{\emph{Vehicular Communications}}  \bibinfo{volume}{35} (\bibinfo{year}{2022}), \bibinfo{pages}{100459}.
\newblock


\bibitem[Evans et~al\mbox{.}(2018)]%
        {introToMPC}
\bibfield{author}{\bibinfo{person}{David Evans}, \bibinfo{person}{Vladimir Kolesnikov}, \bibinfo{person}{Mike Rosulek}, {et~al\mbox{.}}} \bibinfo{year}{2018}\natexlab{}.
\newblock \showarticletitle{A pragmatic introduction to secure multi-party computation}.
\newblock \bibinfo{journal}{\emph{Foundations and Trends{\textregistered} in Privacy and Security}} \bibinfo{volume}{2}, \bibinfo{number}{2-3} (\bibinfo{year}{2018}), \bibinfo{pages}{70--246}.
\newblock


\bibitem[Fiege et~al\mbox{.}(1987)]%
        {zk}
\bibfield{author}{\bibinfo{person}{Uriel Fiege}, \bibinfo{person}{Amos Fiat}, {and} \bibinfo{person}{Adi Shamir}.} \bibinfo{year}{1987}\natexlab{}.
\newblock \showarticletitle{Zero knowledge proofs of identity}. In \bibinfo{booktitle}{\emph{Proceedings of the nineteenth annual ACM symposium on Theory of computing}}. \bibinfo{pages}{210--217}.
\newblock


\bibitem[Goldreich et~al\mbox{.}(2019)]%
        {mpc}
\bibfield{author}{\bibinfo{person}{Oded Goldreich}, \bibinfo{person}{Silvio Micali}, {and} \bibinfo{person}{Avi Wigderson}.} \bibinfo{year}{2019}\natexlab{}.
\newblock \showarticletitle{How to play any mental game, or a completeness theorem for protocols with honest majority}.
\newblock In \bibinfo{booktitle}{\emph{Providing Sound Foundations for Cryptography: On the Work of Shafi Goldwasser and Silvio Micali}}. \bibinfo{pages}{307--328}.
\newblock


\bibitem[Hartmann and Steup(2013)]%
        {hartmann}
\bibfield{author}{\bibinfo{person}{Kim Hartmann} {and} \bibinfo{person}{Christoph Steup}.} \bibinfo{year}{2013}\natexlab{}.
\newblock \showarticletitle{The vulnerability of UAVs to cyber attacks-An approach to the risk assessment}. In \bibinfo{booktitle}{\emph{2013 5th international conference on cyber conflict (CYCON 2013)}}. IEEE, \bibinfo{pages}{1--23}.
\newblock


\bibitem[He et~al\mbox{.}(2023)]%
        {creditScoring}
\bibfield{author}{\bibinfo{person}{Haoran He}, \bibinfo{person}{Zhao Wang}, \bibinfo{person}{Hemant Jain}, \bibinfo{person}{Cuiqing Jiang}, {and} \bibinfo{person}{Shanlin Yang}.} \bibinfo{year}{2023}\natexlab{}.
\newblock \showarticletitle{A privacy-preserving decentralized credit scoring method based on multi-party information}.
\newblock \bibinfo{journal}{\emph{Decision Support Systems}}  \bibinfo{volume}{166} (\bibinfo{year}{2023}), \bibinfo{pages}{113910}.
\newblock


\bibitem[Keller(2020)]%
        {mpspdz}
\bibfield{author}{\bibinfo{person}{Marcel Keller}.} \bibinfo{year}{2020}\natexlab{}.
\newblock \showarticletitle{{MP-SPDZ}: A Versatile Framework for Multi-Party Computation}. In \bibinfo{booktitle}{\emph{Proceedings of the 2020 ACM SIGSAC Conference on Computer and Communications Security}}.
\newblock
\urldef\tempurl%
\url{https://doi.org/10.1145/3372297.3417872}
\showDOI{\tempurl}


\bibitem[Keller et~al\mbox{.}(2016)]%
        {mascot}
\bibfield{author}{\bibinfo{person}{Marcel Keller}, \bibinfo{person}{Emmanuela Orsini}, {and} \bibinfo{person}{Peter Scholl}.} \bibinfo{year}{2016}\natexlab{}.
\newblock \showarticletitle{MASCOT: faster malicious arithmetic secure computation with oblivious transfer}. In \bibinfo{booktitle}{\emph{Proceedings of the 2016 ACM SIGSAC Conference on Computer and Communications Security}}. \bibinfo{pages}{830--842}.
\newblock


\bibitem[Kikuchi et~al\mbox{.}(2018)]%
        {passwordAuth}
\bibfield{author}{\bibinfo{person}{Ryo Kikuchi}, \bibinfo{person}{Koji Chida}, \bibinfo{person}{Dai Ikarashi}, {and} \bibinfo{person}{Koki Hamada}.} \bibinfo{year}{2018}\natexlab{}.
\newblock \showarticletitle{Password-based authentication protocol for secret-sharing-based multiparty computation}.
\newblock \bibinfo{journal}{\emph{IEICE Transactions on Fundamentals of Electronics, Communications and Computer Sciences}} \bibinfo{volume}{101}, \bibinfo{number}{1} (\bibinfo{year}{2018}), \bibinfo{pages}{51--63}.
\newblock


\bibitem[Mishra et~al\mbox{.}(2020)]%
        {searchAndRescue}
\bibfield{author}{\bibinfo{person}{Balmukund Mishra}, \bibinfo{person}{Deepak Garg}, \bibinfo{person}{Pratik Narang}, {and} \bibinfo{person}{Vipul Mishra}.} \bibinfo{year}{2020}\natexlab{}.
\newblock \showarticletitle{Drone-surveillance for search and rescue in natural disaster}.
\newblock \bibinfo{journal}{\emph{Computer Communications}}  \bibinfo{volume}{156} (\bibinfo{year}{2020}), \bibinfo{pages}{1--10}.
\newblock


\bibitem[Ohlmeyer et~al\mbox{.}({[n.\,d.]})]%
        {GPSNav}
\bibfield{author}{\bibinfo{person}{Ernest Ohlmeyer}, \bibinfo{person}{Thomas Pepitone}, \bibinfo{person}{B. Miller}, \bibinfo{person}{D. Malyevac}, \bibinfo{person}{John Bibel}, \bibinfo{person}{Alan Evans}, \bibinfo{person}{Ernest Ohlmeyer}, \bibinfo{person}{Thomas Pepitone}, \bibinfo{person}{B. Miller}, \bibinfo{person}{D. Malyevac}, \bibinfo{person}{John Bibel}, {and} \bibinfo{person}{Alan Evans}.} \bibinfo{year}{[n.\,d.]}\natexlab{}.
\newblock \bibinfo{booktitle}{\emph{GPS-aided navigation system requirements for smart munitions and guided missiles}}.
\newblock


\bibitem[Rabin(2005)]%
        {OT}
\bibfield{author}{\bibinfo{person}{Michael~O Rabin}.} \bibinfo{year}{2005}\natexlab{}.
\newblock \showarticletitle{How to exchange secrets with oblivious transfer}.
\newblock \bibinfo{journal}{\emph{Cryptology ePrint Archive}} (\bibinfo{year}{2005}).
\newblock


\bibitem[Rabta et~al\mbox{.}(2018)]%
        {disRelief}
\bibfield{author}{\bibinfo{person}{Boualem Rabta}, \bibinfo{person}{Christian Wankm{\"u}ller}, {and} \bibinfo{person}{Gerald Reiner}.} \bibinfo{year}{2018}\natexlab{}.
\newblock \showarticletitle{A drone fleet model for last-mile distribution in disaster relief operations}.
\newblock \bibinfo{journal}{\emph{International Journal of Disaster Risk Reduction}}  \bibinfo{volume}{28} (\bibinfo{year}{2018}), \bibinfo{pages}{107--112}.
\newblock


\bibitem[Riazi et~al\mbox{.}(2018)]%
        {chameleon}
\bibfield{author}{\bibinfo{person}{M~Sadegh Riazi}, \bibinfo{person}{Christian Weinert}, \bibinfo{person}{Oleksandr Tkachenko}, \bibinfo{person}{Ebrahim~M Songhori}, \bibinfo{person}{Thomas Schneider}, {and} \bibinfo{person}{Farinaz Koushanfar}.} \bibinfo{year}{2018}\natexlab{}.
\newblock \showarticletitle{Chameleon: A hybrid secure computation framework for machine learning applications}. In \bibinfo{booktitle}{\emph{Proceedings of the 2018 on Asia conference on computer and communications security}}. \bibinfo{pages}{707--721}.
\newblock


\bibitem[Soewito and Marcellinus(2021)]%
        {iot1}
\bibfield{author}{\bibinfo{person}{Benfano Soewito} {and} \bibinfo{person}{Yonathan Marcellinus}.} \bibinfo{year}{2021}\natexlab{}.
\newblock \showarticletitle{IoT security system with modified Zero Knowledge Proof algorithm for authentication}.
\newblock \bibinfo{journal}{\emph{Egyptian Informatics Journal}} \bibinfo{volume}{22}, \bibinfo{number}{3} (\bibinfo{year}{2021}), \bibinfo{pages}{269--276}.
\newblock


\bibitem[Songhori et~al\mbox{.}(2015)]%
        {tg}
\bibfield{author}{\bibinfo{person}{Ebrahim~M Songhori}, \bibinfo{person}{Siam~U Hussain}, \bibinfo{person}{Ahmad-Reza Sadeghi}, \bibinfo{person}{Thomas Schneider}, {and} \bibinfo{person}{Farinaz Koushanfar}.} \bibinfo{year}{2015}\natexlab{}.
\newblock \showarticletitle{Tinygarble: Highly compressed and scalable sequential garbled circuits}. In \bibinfo{booktitle}{\emph{2015 IEEE Symposium on Security and Privacy}}. IEEE, \bibinfo{pages}{411--428}.
\newblock


\bibitem[Svaigen et~al\mbox{.}(2022)]%
        {IODdesign}
\bibfield{author}{\bibinfo{person}{Alisson~R Svaigen}, \bibinfo{person}{Azzedine Boukerche}, \bibinfo{person}{Linnyer~B Ruiz}, {and} \bibinfo{person}{Antonio~AF Loureiro}.} \bibinfo{year}{2022}\natexlab{}.
\newblock \showarticletitle{Design guidelines of the internet of drones location privacy protocols}.
\newblock \bibinfo{journal}{\emph{IEEE Internet of Things Magazine}} \bibinfo{volume}{5}, \bibinfo{number}{2} (\bibinfo{year}{2022}), \bibinfo{pages}{175--180}.
\newblock


\bibitem[Svaigen et~al\mbox{.}(2023)]%
        {jammingAttacks}
\bibfield{author}{\bibinfo{person}{Alisson~R Svaigen}, \bibinfo{person}{Azzedine Boukerche}, \bibinfo{person}{Linnyer~B Ruiz}, {and} \bibinfo{person}{Antonio~AF Loureiro}.} \bibinfo{year}{2023}\natexlab{}.
\newblock \showarticletitle{Trajectory Matters: Impact of Jamming Attacks Over the Drone Path Planning on the Internet of Drones}.
\newblock \bibinfo{journal}{\emph{Ad Hoc Networks}}  \bibinfo{volume}{146} (\bibinfo{year}{2023}), \bibinfo{pages}{103179}.
\newblock


\bibitem[Tedeschi et~al\mbox{.}(2023)]%
        {PPCA}
\bibfield{author}{\bibinfo{person}{Pietro Tedeschi}, \bibinfo{person}{Savio Sciancalepore}, {and} \bibinfo{person}{Roberto Di~Pietro}.} \bibinfo{year}{2023}\natexlab{}.
\newblock \showarticletitle{PPCA - Privacy-Preserving Collision Avoidance for Autonomous Unmanned Aerial Vehicles}.
\newblock \bibinfo{journal}{\emph{IEEE Transactions on Dependable and Secure Computing}} \bibinfo{volume}{20}, \bibinfo{number}{2} (\bibinfo{year}{2023}), \bibinfo{pages}{1541--1558}.
\newblock
\urldef\tempurl%
\url{https://doi.org/10.1109/TDSC.2022.3159837}
\showDOI{\tempurl}


\bibitem[Walshe et~al\mbox{.}(2019)]%
        {iot2}
\bibfield{author}{\bibinfo{person}{Marcus Walshe}, \bibinfo{person}{Gregory Epiphaniou}, \bibinfo{person}{Haider Al-Khateeb}, \bibinfo{person}{Mohammad Hammoudeh}, \bibinfo{person}{Vasilios Katos}, {and} \bibinfo{person}{Ali Dehghantanha}.} \bibinfo{year}{2019}\natexlab{}.
\newblock \showarticletitle{Non-interactive zero knowledge proofs for the authentication of IoT devices in reduced connectivity environments}.
\newblock \bibinfo{journal}{\emph{Ad Hoc Networks}}  \bibinfo{volume}{95} (\bibinfo{year}{2019}), \bibinfo{pages}{101988}.
\newblock


\bibitem[Wang(2022)]%
        {emp}
\bibfield{author}{\bibinfo{person}{Xiao Wang}.} \bibinfo{year}{accessed 2022}\natexlab{}.
\newblock \bibinfo{title}{{EMP-Toolkit}}.
\newblock \bibinfo{howpublished}{\url{https://github.com/emp-toolkit}}.
\newblock


\bibitem[Wazid et~al\mbox{.}(2020)]%
        {blockchain3}
\bibfield{author}{\bibinfo{person}{Mohammad Wazid}, \bibinfo{person}{Basudeb Bera}, \bibinfo{person}{Ankush Mitra}, \bibinfo{person}{Ashok~Kumar Das}, {and} \bibinfo{person}{Rashid Ali}.} \bibinfo{year}{2020}\natexlab{}.
\newblock \showarticletitle{Private blockchain-envisioned security framework for AI-enabled IoT-based drone-aided healthcare services}. In \bibinfo{booktitle}{\emph{Proceedings of the 2nd ACM MobiCom workshop on drone assisted wireless communications for 5G and beyond}}. \bibinfo{pages}{37--42}.
\newblock


\bibitem[Weng et~al\mbox{.}(2021)]%
        {weng2021wolverine}
\bibfield{author}{\bibinfo{person}{Chenkai Weng}, \bibinfo{person}{Kang Yang}, \bibinfo{person}{Jonathan Katz}, {and} \bibinfo{person}{Xiao Wang}.} \bibinfo{year}{2021}\natexlab{}.
\newblock \showarticletitle{Wolverine: fast, scalable, and communication-efficient zero-knowledge proofs for boolean and arithmetic circuits}. In \bibinfo{booktitle}{\emph{2021 IEEE Symposium on Security and Privacy (SP)}}. IEEE, \bibinfo{pages}{1074--1091}.
\newblock


\bibitem[Yao(1982)]%
        {2pc}
\bibfield{author}{\bibinfo{person}{Andrew~C. Yao}.} \bibinfo{year}{1982}\natexlab{}.
\newblock \showarticletitle{Protocols for secure computations}. In \bibinfo{booktitle}{\emph{23rd Annual Symposium on Foundations of Computer Science (sfcs 1982)}}. \bibinfo{pages}{160--164}.
\newblock
\urldef\tempurl%
\url{https://doi.org/10.1109/SFCS.1982.38}
\showDOI{\tempurl}


\bibitem[Yue et~al\mbox{.}(2022)]%
        {interceptComm}
\bibfield{author}{\bibinfo{person}{Pingyue Yue}, \bibinfo{person}{Jianping An}, \bibinfo{person}{Jiankang Zhang}, \bibinfo{person}{Gaofeng Pan}, \bibinfo{person}{Shuai Wang}, \bibinfo{person}{Pei Xiao}, {and} \bibinfo{person}{Lajos Hanzo}.} \bibinfo{year}{2022}\natexlab{}.
\newblock \showarticletitle{On the security of LEO satellite communication systems: Vulnerabilities, countermeasures, and future trends}.
\newblock \bibinfo{journal}{\emph{arXiv preprint arXiv:2201.03063}} (\bibinfo{year}{2022}).
\newblock


\end{thebibliography}

\end{document}